\documentclass[twocolumn,english,aps,prb]{revtex4-1}
\usepackage[T1]{fontenc}
\usepackage[latin9]{inputenc}
\usepackage{amsmath}
\usepackage{amssymb}
\usepackage{graphicx}
\usepackage{wasysym}
\usepackage{esint}

\makeatletter
\usepackage{babel}
\usepackage{xcolor}

\makeatother

\usepackage{babel}
\begin{document}

\title{Gapless regime in the charge density wave phase of the finite dimensional
Falicov-Kimball model}

\author{Martin \v{Z}onda, Junichi Okamoto and Michael Thoss}

\address{Institute of Physics, Albert-Ludwig University of Freiburg, Hermann-Herder-Strasse
3, 791 04 Freiburg, Germany}
\begin{abstract}
The ground-state density of states of the half-filled Falicov-Kimball
model contains a charge-density-wave gap. At finite temperature, this
gap is not immediately closed, but is rather filled in by subgap states.
For a specific combination of parameters, this leads to a stable phase
where the system is in an ordered charge-density-wave phase, but there
is high density of states at the Fermi level. We show that this property
can be, in finite dimensions, traced to a crossing of sharp states resulting
from the single particle excitations of the localized subsystem. The
analysis of the inverse participation ratio points to a strong localization
in the discussed regime. However, the pronounced subgap density of
states can still lead to a notable increase of charge transport through
a finite size system. We show this by focusing on the transmission
in heterostructures where a Falicov-Kimball system is sandwiched between
two metallic leads. 
\end{abstract}
\maketitle

\section{Introduction}

The Falicov-Kimball model (FKM) \cite{FKM1969} is one of the simplest
models for the description of correlated electrons on a lattice. Over
time, it has become a standard tool for the investigation of various
phenomena including crystallization \cite{Kennedy_1986,Jedrzejewski1989,Gruber1994},
metal-insulator and valence transitions 
\cite{Plischke1972,Michielsen1994,Farky1995a,Farky1995b,Portengen1996,Czycholl1999,Byczuk2005,Lemanski2017,Haldar2017b,Farky2019,Nasu2019,Haldar2019},
inhomogeneous charge and spin orderings ~\cite{Lemberger1992,Freericks_seg_1999,LemanskiPRL2002,Freericks2002,Lemanski2004,Tran2006,Cencarikova2011,Zonda2012,Debski2016,Kapcia2017},
nonlocal correlations \cite{Schiller1999,Hetler2000,Ribic2016,Ribic2017},
ferroelectricity \cite{BatistaPRL2002,FarkyFero2002,FarkyFero2008,Schneider2008,Golosov2013},
mixtures of heavy and light cold atoms in optical lattices \cite{MaskaPRL2008,Iskin2009,MaskaPRA2011,HuMaska2015},
transport through layered systems \cite{Freericks2001,Freericks2004,FreericksBook2006,Hale2012,Kaneko2013,Zonda2018}
or different nonequilibrium phenomena \cite{FreericksPRL2006,Turkowski2007,EcksteinPRL2008,Eckstein2009,Herrmann2016,Haldar2016,Haldar2017,Smith2017,Qin2018,Herrmann2018}.

Its biggest advantage over the paradigmatic Hubbard model \cite{Hubbard63}
lies in the fact that it is accessible by exact methods. It is exactly
solvable in the limit of infinite dimensions (infinite coordination
number) by means of dynamical mean field theory (DMFT) \cite{Brandt1989,Brandt1990,Janis1991,FreericksRMP2003}
and it can be addressed by an exact, sign-problem-free Monte Carlo
(MC) method \cite{MaskaPRB2006,ZondaSSC2009,Zonda2012,Huang2017}
in finite dimensions. Both methods take advantage of the fact that
the FKM combines quantum and classical degrees of freedom.

Despite the simplicity and accessibility of this model, its research
continues to offer new and often surprising results. This is true
even for its simplest spin-less version at the particle-hole symmetric
point. For example, nonequilibrium DMFT and cluster approximation
method studies showed that its quantum subsystem does not thermalize
after an interaction quench \cite{EcksteinPRL2008,Herrmann2016,Herrmann2018};
simple generalizations of the FKM can be used to study the interplay
of topology and interactions at finite temperatures \cite{Goncalves2019}
or fractionalization of particles into charge and spin objects \cite{Tran2019};
it was used to derive universal features of the critical metal-insulator
transition that are transferable to other Hubbard-like models \cite{Janis2014,Haldar2017b},
utilized in studies of different quasiparticles \cite{Prosko2017,Kauch2019};
and, as discussed below, the phase diagram of the model is still in
question as well.

The Hamiltonian of the spinless FKM at half filling reads 
\begin{eqnarray}
H_{\mathrm{FK}} & = & -t\sum_{\left\langle \boldsymbol{l},\boldsymbol{l}'\right\rangle }\left(d_{\boldsymbol{l}}^{\dagger}d_{\boldsymbol{l}'}^{\phantom{\dagger}}+d_{\boldsymbol{l}'}^{\dagger}d_{\boldsymbol{l}}^{\phantom{\dagger}}\right)\nonumber \\
 &  & +U\sum_{\boldsymbol{l}}\left(f_{\boldsymbol{l}}^{\dagger}f_{\boldsymbol{l}}^{\phantom{\dagger}}-\frac{1}{2}\right)\left(d_{\boldsymbol{l}}^{\dagger}d_{\boldsymbol{l}}^{\phantom{\dagger}}-\frac{1}{2}\right)\label{eq:Model}
\end{eqnarray}
where the first term describes nearest-neighbor hopping of spinless
electrons on a lattice. The second term represents a Coulomb-like
local interaction between the localized $f$ particle and itinerant
$d$ electron on the same lattice site. The terms with factor $-\frac{1}{2}$
secure the half-filling conditions $N_{f}+N_{d}=L/2$ for chemical
potential $\mu=0$. Here $N_{f(d)}$ is the total number of $f$ ($d$)
particles and $L$ is the total number of lattice points.

\begin{figure}
\includegraphics[width=0.75\columnwidth]{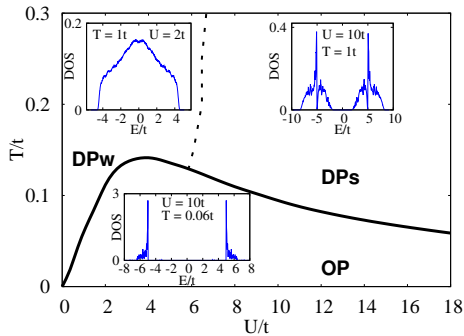}\caption{Simplified phase diagram of the spinless FKM on a square 2D lattice
with ordered CDW phase (OP) and disordered phases in weak (DPw) and
strong (DPs) interaction regimes. The insets illustrate typical $d$-electron
DOS in the respective phases. \label{fig:PhD}}
\end{figure}

The phase diagram of this model in finite as well as infinite dimensions
contains three main phases (see Fig.~\ref{fig:PhD}): an ordered
charge density wave (CDW) phase (OP) that exists at low temperatures
\cite{Brandt1989,Brandt1990,Brandt1991,ChenPRB2003,ZondaSSC2009},
a gapless disordered phase for weak interaction $U$ and high temperatures
(DPw), and a gapped disordered phase for strong interaction $U$ and
high temperatures (DPs) \cite{GruberHPA1996,FreericksRMP2003}.

However, this is not a complete picture. A recent study of the model
on a two-dimensional ($D=2$) lattice \cite{Antipov2016} showed
that in the thermodynamic limit DPw exhibits Anderson localization
which destabilizes the metallic-like phase reported in older works
focused on relatively small lattice sizes \cite{MaskaPRB2006,ZondaSSC2009}.
Therefore, all three phases are insulating in the thermodynamics limit.
Nevertheless, for any finite system size, there is a crossover from
an Anderson localized insulating phase at intermediate $U$ through
a weakly localized regime, with the above mentioned metallic-like
character, to a Fermi gas at $U=0$. In addition, a series of papers
proved that in infinite dimensions ($D\rightarrow\infty$) there is
a stable ordered CDW phase without a gap at the Fermi level \cite{Hassan2007,MatveevPRB2008,Lemanski2014,LemanskiAPP_2016,Kapcia2019}.

The gapless CDW phase in the infinite dimensional FKM is related to
the existence of narrow bands in the density of states (DOS) that
are formed inside the CDW gap at finite temperatures. These subgap
bands come in pairs placed symmetrically around the center of the gap
and merge for a finite range of interaction strengths $U$ and temperatures.
The resulting merged single subgap band is centered around the Fermi
level and, consequently, there is no gap at the Fermi level. Lema\'{n}ski
argued that this merging is related to the inversion of the subgap
bands belonging to two sublattices of a bipartite lattice at critical
interaction $U_{c}$ \cite{LemanskiAPP_2016}. Here a sublattice A
constitutes such lattice points that all their nearest neighbors
belong to the sublattice B and vice versa (they are alternating).
The DOS calculated for each sublattice separately contains both subgap
bands placed symmetrically around the Fermi energy. However, they
differ in width and height. This property is related to the CDW ordering
because the average $f$-electron occupancy differs for the sublattices
as was discussed in various studies \cite{Hassan2007,MatveevPRB2008,Lemanski2014}.
What is interesting is that at some critical $U_{c}$ the qualitatively
different subgap bands belonging to different sublattices flip positions.

The open question is if there is an analog of such a band crossing
in finite dimensions as well. The subgap states, respective bands,
had already been discussed in both $D=2$ and $D=3$ \cite{MaskaPRB2006,Tran2006,ZondaSSC2009}.
However, a systematic study focused on the region around $U_{c}$
is missing. The present paper has the aim to fill this gap. We show
that there is indeed a clear crossing of distinct subgap states in
finite dimensions. Moreover, we reveal the underlying mechanism of
the crossing by focusing on the single particle excitations from the
CDW ground state. We show that these excitations significantly influence
the density of states up to surprisingly high temperatures approaching
$T_{c}$ of the order-disorder transition. We also demonstrate that,
despite the presence of strong localization, the crossing can support
charge transmission through a finite-sized system in the gapped regime.
This is done by addressing a heterostructure where the finite system
modeled by the FKM is sandwiched between two metallic leads. We mostly
focus on the $D=2$ case, but $D=1$ and $D=3$ are addressed as well.

The rest of the paper is organized as follows. In Sec.~\ref{sec:methods}
we outline the methodology for addressing the thermodynamic properties
of the FKM and introduce the model of the heterostructure as well
as the method for studying charge transport in it. The main results
are presented in Sec.~\ref{sec:Results}, where we first analyze
the origin and properties of the sharp features of the subgap DOS
in Sec.~\ref{subsec:DOS} and then show in Sec.~\ref{subsec:Transport}
how these affect the charge transport through finite FKM coupled to
metallic leads. Section~\ref{sec:Conclusions} concludes with a summary.
In the Appendices we show some analytical results on the existence
of the gap and the positions of sharp subgap states.

\section{Methods\label{sec:methods} }

\subsection{Thermodynamic properties}

The $f$ particles in the FKM represent classical degrees of freedom.
This can be seen from the fact that the $f$-particle occupation numbers
$f_{\boldsymbol{l}}^{\dagger}f_{\boldsymbol{l}}^{\phantom{\dagger}}$
commute with the entire Hamiltonian in Eq.~(\ref{eq:Model}). This
allows us to replace any operator $f_{\boldsymbol{l}}^{\dagger}f_{\boldsymbol{l}}^{\phantom{\dagger}}$
by its eigenvalues $w_{\boldsymbol{l}}=0$ or $w_{\boldsymbol{l}}=1$
and write a partial Hamiltonian for a particular classical configuration
$w$. After neglecting the constant energy shift $-U/4$, the Hamiltonian
in Eq.~\eqref{eq:Model} reads for a chosen configuration $w$ 
\begin{equation}
H_{w}=\sum_{\boldsymbol{l},\boldsymbol{l}'}h_{\boldsymbol{l\,l}'}d_{\boldsymbol{l}}^{\dagger}d_{\boldsymbol{l}'}^{\phantom{\dagger}}-\frac{U}{2}N_{f}=\sum_{\alpha}\varepsilon_{\alpha}\tilde{d}_{\alpha}^{\dagger}\tilde{d}_{\alpha}^{\phantom{\dagger}}-\frac{U}{2}N_{f}.\label{eq:H1tr}
\end{equation}
Thereby, $\varepsilon_{j}$ are the eigenvalues of the matrix with
elements $h_{\boldsymbol{l\,l}'}=Uw_{\boldsymbol{l}}\delta_{\boldsymbol{l\,l}'}-t\delta_{\left\langle \boldsymbol{l,l}'\right\rangle }$,
where $\delta_{\left\langle \boldsymbol{l,l}'\right\rangle }=1$ when
the lattice positions represented by vectors $\boldsymbol{l}$ and
$\boldsymbol{l}'$ are the nearest neighbors and zero otherwise. The
ground-state configuration for any bipartite lattice at the particle-hole
symmetric point is the checkerboard ordering of $f$ particles \cite{Kennedy_1986,GruberHPA1996}.
The corresponding configuration $w$ can be written as $w_{\mathbf{l}}=(1+e^{i\boldsymbol{\pi}\boldsymbol{l}})/2$
for any hypercubic lattice. It is easy to show (see Appendix~A) that
such an effective potential opens a gap of the width $U$ in the band
structure which is centered around the Fermi energy.

An advantage of the Falicov-Kimball model is that the mean values
of any $d$-electron operator $\hat{O}$ can be written in the form
\begin{equation}
\left\langle \hat{O}\right\rangle =\mathrm{Tr}_{w}\left\langle \hat{O}\right\rangle _{d}\equiv\frac{1}{Z}\sum_{w}e^{-\beta F\left(w\right)}\left\langle \hat{O}\right\rangle _{d},\label{eq:Aver}
\end{equation}
where 
\begin{equation}
F\left(w\right)=-\frac{1}{\beta}\sum_{\alpha}\ln\left[1+e^{-\beta\varepsilon_{\alpha}}\right]-\frac{U}{2}N_{f},\label{eq:Free}
\end{equation}
with $Z=\sum_{w}e^{-\beta F\left(w\right)}$ being the partition function
(we assume $\mu=0$). Here $\left\langle .\right\rangle _{d}$ is
the trace over the $d$-electron subsystem for fixed $w$ \cite{MaskaPRB2006}.
As this is a single-particle problem, the trace can be calculated
effectively using exact numerical diagonalization. The sum over configurations
$w$ can then be calculated using a Metropolis algorithm based Monte
Carlo method \cite{MaskaPRB2006,ZondaSSC2009,Zonda2009,Zonda2012,Huang2017}.

The quantities that we are mostly interested in are the normalized
DOS defined as 
\begin{equation}
\mathrm{DOS}(\varepsilon)=\frac{1}{L}\mathrm{Tr}_{w}\sum_{\alpha}\delta\left(\varepsilon-\varepsilon_{\alpha}\right)
\end{equation}
 where $\mathrm{Tr}_{w}\equiv\frac{1}{Z}\sum_{w}e^{-\beta F\left(w\right)}$
and averaged inverse participation ratio (IPR) 
\begin{equation}
\mathrm{IPR}(\varepsilon)=\mathrm{Tr}_{w}\frac{\sum_{i}\mathrm{DOS}_{i}(\varepsilon,w)^{2}}{\mathrm{DOS}(\varepsilon,w)^{2}},\label{eq:IPR}
\end{equation}
where $\mathrm{DOS}_{i}(\varepsilon,w)=\sum_{\alpha}\delta\left(\varepsilon-\varepsilon_{\alpha}\right)\mathcal{U}_{i\alpha}\mathcal{U}_{\alpha i}^{\dagger}/L$
is the local DOS and the matrix $\mathcal{U}$ consists of the eigenvectors
belonging to eigenvalues $\varepsilon_{\alpha}$ of the matrix $\boldsymbol{h}$
from Eq.~\eqref{eq:H1tr} calculated for a particular configuration
$w$ and arranged in columns. The matrix $\mathcal{U}$ can be evaluated
numerically for a finite system and it can be chosen to be real.

The finite size scaling of the IPR can be used for studying localization
of the itinerant electrons in the system \cite{Evers2008,Murphy2011,Perera2018}.
The $\mathrm{IPR}$ scales as $1/L$ for a completely itinerant system
states and converges to a finite value with increasing $L$ for strongly
localized states. In the case of perfect spacial localization to a
single lattice point, the IPR converges to one. Note, that because
of the finite size of our lattices, we use a Gaussian broadening of
the $\delta$-functions, $\delta\left(\varepsilon-\varepsilon_{\alpha}\right)\approx\exp[-(\varepsilon-\varepsilon_{\alpha})^{2}/(2b^{2})]/(b\sqrt{2\pi})$.
In most of presented cases, we set the broadening constant to $b=0.002t$.
This small value is a compromise between the effort to suppress the
influence of the artificial broadening on our results (especially
IPR) and the preservation of the stability of the calculations of
the IPR for a broad range of temperatures and lattice sizes. We discuss
the influence of the Gaussian broadening on our results in detail
below.

\subsection{Heterostructure}

\begin{figure}
\includegraphics[width=1\columnwidth]{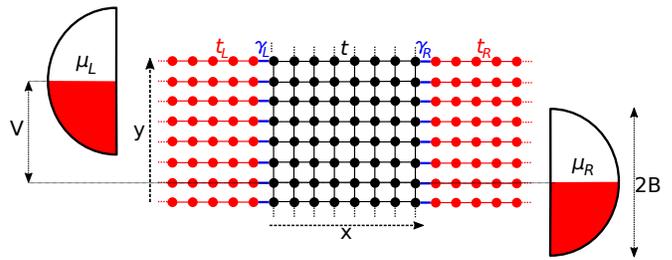}\caption{Schematic picture of the heterostructure. The black part represents
the two-dimensional FKM system with nearest neighbor hopping $t$.
The red parts are noninteracting leads with hopping $t_{L,R}$ and
the hybridization interaction with hopping $\gamma_{L,R}$ is indicated
in blue. The leads are characterized by elliptic surface DOSs. The
voltage drop $V$ is introduced by mutual shift of $\epsilon_{L,R}$
where the condition $\mu_{L,R}=\epsilon_{L,R}$ is used to keep the
bands half filled at any applied voltage. \label{fig:schema}}
\end{figure}

Besides studying an isolated FKM, we also address a heterostructure
$H_{\mathrm{h}}=H_{\mathrm{FK}}+\sum_{l=L,R}H_{\mathrm{lead}}^{l}+H_{\mathrm{hyb}}^{l}$,
where a two dimensional FK system is sandwiched between two metallic
leads as illustrated in Fig.~\ref{fig:schema}. The central system
$H_{\mathrm{FK}}$ is finite in $x$ but in principle infinite in the
$y$ direction (modeled by periodic boundary conditions). The leads
and hybridization terms read 
\begin{eqnarray}
H_{\mathrm{lead}}^{l} & = & -t_{l}\sum_{\left\langle m,n\right\rangle }\left(c_{l,m}^{\dagger}c_{l,n}^{\phantom{\dagger}}+c_{l,n}^{\phantom{\dagger}}c_{l,m}^{\dagger}\right)+\epsilon_{l}\sum_{n}c_{l,n}^{\dagger}c_{l,n}^{\phantom{\dagger}},\nonumber \\
H_{\mathrm{hyb}}^{l} & = & -\gamma_{l}\sum_{\left\langle i,n\right\rangle }\left(c_{l,n}^{\dagger}d_{i}^{\phantom{\dagger}}+d_{i}^{\dagger}c_{l,n}^{\phantom{\dagger}}\right),\label{eq:Lead}
\end{eqnarray}
where $t_{l}$ is the hopping for lead $l=L,R$ , $\epsilon$ represents
an energy shift of the lead, and $\gamma_{l}$ is hopping parameter
between the system and the lead $l$. 

We have two main motivations for addressing this more complex setup.
First, the broadening of the system states is in the case of the heterostructure
provided naturally by the coupling $\gamma_{l}$ to the semi-infinite
leads, which allows to test the results obtained for isolated system
potentially influenced by an artificial broadening of the $\delta$ functions.
Second, we want to reveal how the existence of the finite DOS in the
CDW gap influences the transport properties of the model.

The properties of the heterostructure can be effectively addressed
by a combination of a sign-problem free Monte-Carlo with the nonequilibrium
Green's function technique approach \cite{Zonda2018}. We assume in
our analysis that the central FK system was in the distant past decoupled
from the leads and that both system and leads had been in thermal
equilibrium. The occupation numbers of the $f$ electrons are integrals
of motion, therefore, their distribution can be calculated for the
isolated system as it will not change after coupling to the leads.
Here, in contrast to the previous subsection, we assume open boundary
conditions of the central system at the system-lead interface. Test
calculations show, that if system is large enough ($L_{x}>10$) the
influence of the boundary conditions on the $f$-electron distribution
is negligible. We further assume, that the semi-infinite leads are
unaffected by the system and are modeled by parallel chains coupled
to the central system as shown Fig.~\ref{fig:schema}. Therefore,
the leads can be characterized by their surface density of states

\[
\rho_{l}\left(\varepsilon\right)=\frac{2}{\pi B^{2}}\sqrt{B^{2}-\left(\varepsilon-\epsilon_{l}\right)^{2}},
\]
with band half-width $B=2t_{l}$ centered around the band energy shift
$\epsilon_{l}$ from Eqs.~\eqref{eq:Lead}. This allow us to write
an exact formal solution for the Green's function of the heterostructure
for a particular configuration $w$ (for details, see Refs.~\cite{Zonda2018,Jauho1994}):
\begin{eqnarray}
\mathbf{G}^{r,a}\left(\varepsilon\right) & = & \boldsymbol{g}^{r,a}\left(\varepsilon\right)+\boldsymbol{g}^{r,a}\left(\varepsilon\right)\mathbf{\Sigma}^{r,a}\left(\varepsilon\right)\mathbf{G}^{r,a}\left(\varepsilon\right),\label{eq:Gra}\\
\mathbf{G}^{<}\left(\varepsilon\right) & = & \mathbf{G}^{r}\left(\varepsilon\right)\mathbf{\Sigma}^{<}\left(\varepsilon\right)\mathbf{G}^{a}\left(\varepsilon\right).
\end{eqnarray}
Here, $\mathbf{G}^{r\,(a)}$ is the retarded (advanced) Green's function
of the coupled system, $\mathbf{G}^{<}$ is the lesser Green's function
of the coupled system, and $\boldsymbol{g}^{r\,(a)}\left(\varepsilon\right)$
is the retarded (advance) Green's function of the bare system with
components: 
\begin{equation}
g_{\alpha\beta}^{r,a}\left(\varepsilon\right)=\frac{\delta_{\alpha\beta}}{\varepsilon-\varepsilon_{\alpha}\pm i0}.
\end{equation}
The total tunneling self-energies $\mathbf{\Sigma}^{r,a,<}=\mathbf{\Sigma}_{L}^{r,a,<}+\mathbf{\Sigma}_{R}^{r,a,<}$
have the components 
\begin{eqnarray}
\Sigma_{l,\alpha\beta}^{r,a}(\varepsilon) & = & \Lambda_{l,\alpha\beta}(\varepsilon)\pm\frac{i}{2}\Gamma_{l,\alpha\beta}(\varepsilon),\nonumber \\
\Sigma_{l,\alpha\beta}^{<}(\varepsilon) & = & i\Gamma_{l,\alpha\beta}(\varepsilon)\,f_{l}(\varepsilon-\mu_{l}),\nonumber \\
\Gamma_{l,\alpha\beta}(\varepsilon) & = & 2\pi\gamma^{2}\mathcal{U}_{\alpha\beta}^{\{s^{l}\}}\rho_{l}(\varepsilon),\label{eq:Gamma}\\
\Lambda_{l,\alpha\beta}(\varepsilon) & = & \begin{cases}
\frac{2\gamma^{2}}{B^{2}}\mathcal{U}_{\alpha\beta}^{\{s^{l}\}}\left(\varepsilon-\epsilon_{l}\right)\:\textrm{for}\:\left|\varepsilon-\epsilon_{l}\right|<B\\
\frac{2\gamma^{2}}{B^{2}}\mathcal{U}_{\alpha\beta}^{\{s^{l}\}}\left[\left(\varepsilon-\epsilon_{l}\right)\mp\sqrt{\left(\varepsilon-\epsilon_{l}\right)^{2}-B^{2}}\right]\\
\qquad\textrm{for}\:\left(\varepsilon-\epsilon_{l}\right)\gtrless\pm B
\end{cases}\nonumber \\
\mathcal{U}_{\alpha\beta}^{\{s^{l}\}} & = & \sum_{\begin{array}{c}
i\in\{s^{l}\}\end{array}}\mathcal{U}_{\beta i}^{\dagger}\mathcal{U}_{i\alpha},\nonumber 
\end{eqnarray}
where $\{s^{L,R}\}$ are the sets of system lattice positions at the
left and right interfaces, $f_{l}(\varepsilon)$ is the Fermi function, and
$\mu_{l=L,R}$ is the chemical potential of the leads. 

The transmission function, which contains the most detailed information
on charge transport, has for a specific $w$ a compact form \cite{Haug-Jauho},
\begin{equation}
\Theta^{w}\left(\varepsilon\right)=\mathrm{Tr_{d}\,}\left[\mathbf{\Gamma}_{L}\left(\varepsilon\right)\mathbf{G}^{r}\left(\varepsilon\right)\mathbf{\Gamma}_{R}\left(\varepsilon\right)\mathbf{G}^{a}\left(\varepsilon\right)\right],\label{eq:Transmission}
\end{equation}
where the trace goes over the $d$-electron subsystem. Its total mean
value is obtained by a trace over the $f$-electron subsystem which
is done by the MC method. Similarly, the local density of states of
a coupled system (heterostructure) for a specific $w$ can be calculated
as 
\begin{eqnarray}
\mathrm{LDOSh_{\mathit{i}}(\varepsilon,\mathit{w})} & = & \frac{i}{2\pi L}\sum_{\alpha,\beta}\mathcal{U}_{i\alpha}\mathcal{U}_{\beta i}^{\dagger}G_{\alpha\beta}^{r}\left(\varepsilon\right)\label{eq:LDOS}\\
 &  & -\sum_{\alpha,\beta}\mathcal{U}_{\alpha i}^{\dagger}\mathcal{U}_{i\beta}G_{\alpha\beta}^{a}\left(\varepsilon\right),\nonumber 
\end{eqnarray}
and it allows us to define an averaged generalized inverse participation
ratio (gIPR) 
\begin{equation}
\mathrm{gIPR}(E)=\mathrm{Tr}_{w}\frac{\sum_{i}\mathrm{LDOSh_{\mathit{i}}^{2}\mathit{(\varepsilon,w)}}}{\mathrm{DOSh}^{2}(\varepsilon,w)},\label{eq:gIPR}
\end{equation}
where $\mathrm{DOSh}(\varepsilon,w)=\mathrm{Tr}_{d}i\left[\mathbf{G}^{r}\left(\varepsilon\right)-\mathbf{G}^{a}\left(\varepsilon\right)\right]/2\pi L$.

In the following, we set $\mu_{L}=\epsilon_{L}$ and $\mu_{R}=\epsilon_{R}$,
which corresponds to half-filled lead bands and we introduce a finite
voltage drop as $V=\mu_{L}-\mu_{R}$ with antisymmetric condition
$\mu_{L}=-\mu_{R}$. In equilibrium, we set $\mu_{L}=\mu_{R}=0$ and
assume that the electrostatic potential (set to zero) of the central
system is uninfluenced by the leads. This fixes the half-filling condition
for the central system. We focus on equivalent leads $\gamma=\gamma_{L}=\gamma_{R}$
with a broad band half-width $B=10t$.

\section{Results\label{sec:Results}}

\subsection{Subgap density of states\label{subsec:DOS}}

\begin{figure}
\textemdash \includegraphics[width=1\columnwidth]{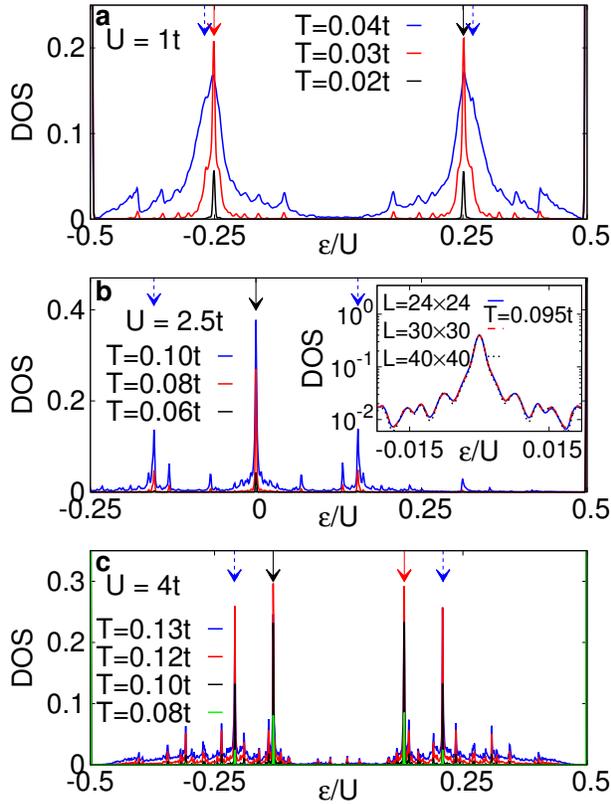}\caption{Details of the subgap density of states ($\left|\varepsilon\right|<U/2$)
for three values of of $U$ and various values of temperature. All
data have been calculated on a two-dimensional square lattice with
size $L=24\times24$ and periodic boundary conditions. The sharp states,
calculated for particular configurations $w$ in the sampling process,
have been broadened by a Gaussian broadening (see text) with $b=0.002t$.
The arrows signal the positions of the bound states belonging to single
$f$-electron excitations discussed in the text. Their color coding
is the same as in Fig.~\ref{fig:exc}(b). The inset in panel (b)
is the detail of the DOS close to the Fermi level calculated for $L=24\times24$,
$30\times30$ and $b=0.002t$ at $T=0.095t$. \label{fig:DOS}}
\end{figure}

The typical subgap bands, calculated for $D=2$ at temperatures in
the vicinity of the CDW phase transition \cite{MaskaPRB2006,ZondaSSC2009,Antipov2016},
resemble the exact DMFT results calculated for infinite dimensions
\cite{Hassan2007,MatveevPRB2008,Lemanski2014,LemanskiAPP_2016}. However,
in finite dimensions the subgap bands reduce with the decreasing temperature
into sharp features pinned mostly to few distinct energies. This is
illustrated in Fig.~\ref{fig:DOS}, where we show the subgap ($\left|\varepsilon\right|<U/2$)
low-temperature DOS calculated for $L=24\times24$, three values
of $U$, and various temperatures below $T_{c}$.

The character of the subgap DOS changes with $U$. For $U=1t$, there
are two pronounced maxima placed symmetrically at $\varepsilon\sim\pm0.25U$;
for $U=2.5t$, a sharp maximum is centered around $\varepsilon=0$
and is accompanied by two sharp features of approximately a third of
its height, which are placed at $\varepsilon\sim\pm0.15U$; for $U=4t$,
four sharp local maxima of comparable heights exist at $\varepsilon\sim\pm0.13U$
and $\varepsilon\sim\pm0.21U$. Considering their positions, the subgap
maxima have the same qualitative dependence on $U$ that was described
for the subgap bands for $D\rightarrow\infty$ \cite{Hassan2007,MatveevPRB2008,Lemanski2014,LemanskiAPP_2016}.
Namely, the maxima approach each other with increasing $U$ until
they merge at some critical $U_{c}$ and then, above it, draw apart.

The position of these most pronounced distinct local maxima does not
depend on the temperature and, in contrast to other states, their
weight is not completely negligible even for a very low temperatures.

\begin{figure}
\includegraphics[width=1\columnwidth]{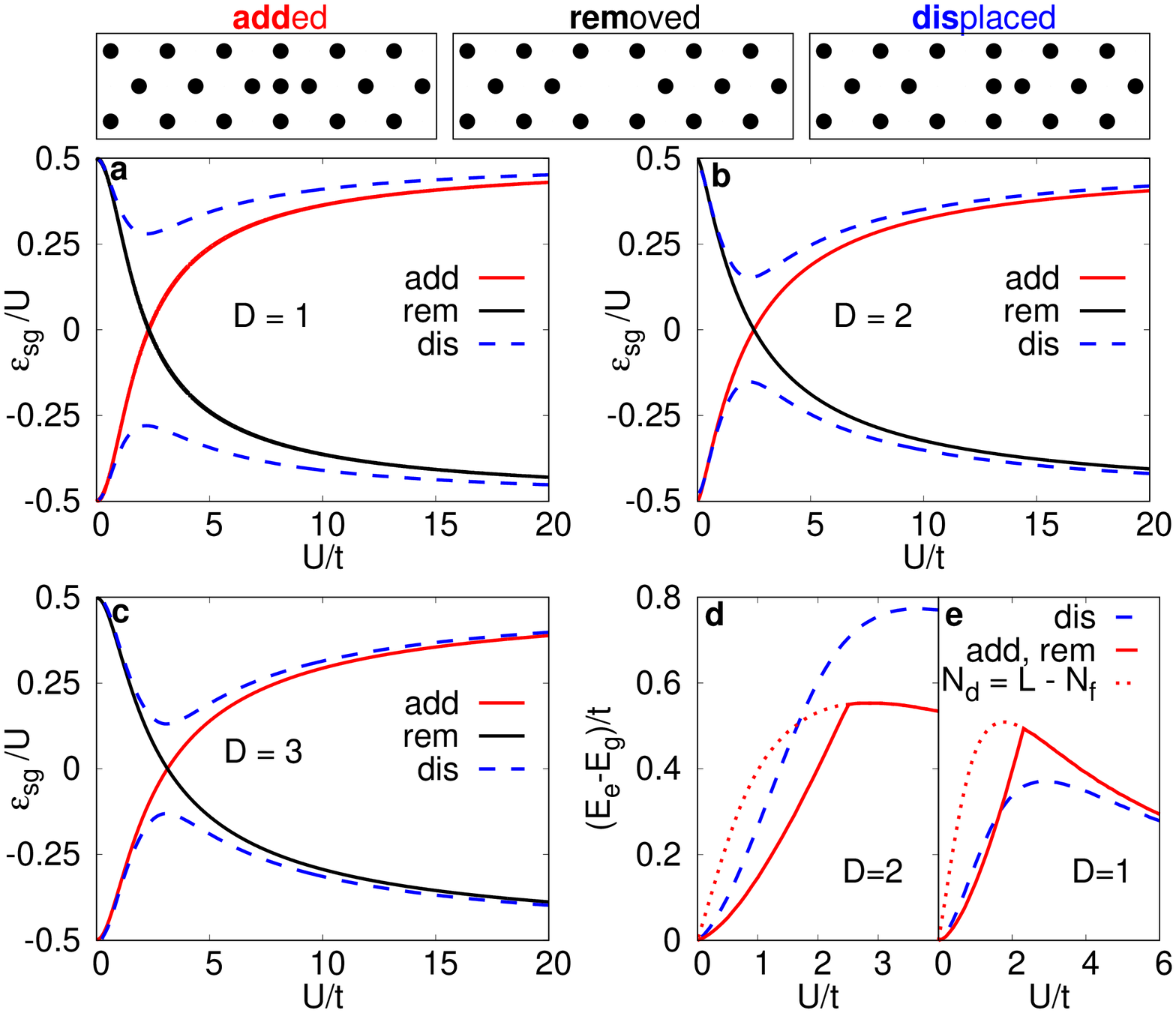}\caption{a-c) Position of the subgap states for disrupted checkerboard ordering
of $f$ electrons in different dimensions. Three single $f$-electron
excitations are considered. Namely, adding one additional $f$ electron
to an unoccupied lattice point (add), removing one $f$ electron (rem)
and displacing one electron from its position to a nearest neighboring
unoccupied point (dis) as illustrated for $D=2$ case in the top three
panels. The results have been obtained using the analytical and semi-analytical
formulas from Appendix~B and represent thermodynamic limit solutions
($L\rightarrow\infty$). d-e) Difference between the minimal energy
of the disrupted and perfect checkerboard ordering of $f$ electrons.
The energy differences for add and rem cases are identical. The dotted
line represents the energy of the system where the half-filling $N_{f}+N_{d}=L$
is enforced. Solid red line breaks this condition below $U_{c}$ by
one particle $N_{f}+N_{d}=L\pm1$. \label{fig:exc}}
\end{figure}

Because the $d$-electron DOS is dictated by the distribution of $f$
particles, one can expect that the origin of these distinct maxima
must be in some states reflecting the low energy disruptions of the
ground-state checkerboard ordering. To analyze this conjecture we
focus on three typical single $f$-electron excitations from the checkerboard
ordering illustrated in the top panels of Fig.~\ref{fig:exc}. They
represent an addition of one $f$ electron to the otherwise perfect
checkerboard ordering; a removal of one $f$ electron and, finally,
a displacement of a single $f$ electron to the nearest unoccupied
lattice point. The actual spacial position of these three defects
does not play a role, as we are assuming periodic boundary conditions.
Note that because $f$ electrons can be also interpreted as ions,
the studied defects can be seen as an equivalent of typical lattice
defects such as vacancies (Schottky defects) or interstitial defects
(Frenkel defects).

Because the above disruptions of the checkerboard pattern play a role
of classical single impurities, they lead to sharp bound states inside
of the CDW gap \cite{Economou2006}. We plot the normalized position
of these states (i.e., their calculated eigenenergies) in Fig.~\ref{fig:exc}
as a function of $U$ in all three realistic dimensions (for details
and analytical results, see Appendix~B). The positions of the the
main peaks in the finite temperature subgap DOS plotted in Fig.~\ref{fig:DOS}
coincide with the eigenenergies at the same $U$ and $D=2$ shown
in Fig.~\ref{fig:exc}(b) (they are also signaled by the arrows in
Fig.~\ref{fig:DOS}). The highest peaks in Fig.~\ref{fig:DOS} reflect
an addition or removal of an $f$ electron and the second in order
reflect a single $f$-electron displacement. Therefore we can conclude
that these simple $f$-electron excitations are the main underlaying
mechanism for the stable finite-temperature subgap anomalies.

The results show that displacing an $f$ electron leads to two subgap
bound states placed symmetrically around $\varepsilon=0$. Adding
or removing an $f$ electron leads to a single subgap state. These
are related by $\varepsilon_{\mathrm{add}}=-\varepsilon_{\mathrm{rem}}$
and cross each other at critical $U_{c}$ which depends on the dimension.
For the one-dimensional case we get $U_{c}=4/\sqrt{3}$, for $D=2$ it
is $U_{c}\cong2.5$, and $D=3$ leads to $U_{c}\cong3.18t$ (see Appendix~B)\footnote{Note, that this crossing resembles the situation in other gaped systems
with bound states, e.g. superconducting quantum dots \cite{Balatsky2006,Meden2019}.}. This means that similarly to the infinite dimensional case \cite{LemanskiAPP_2016},
the main subgap features exchange roles at $U_{c}$.

Another qualitative change that takes place at $U_{c}$ can be seen
from the difference between the minimal energy of our disrupted configurations
($E_{e}$) and the real ground-state energy of the checkerboard ordering
($E_{g}$) plotted in Fig.~\ref{fig:exc}(d). There is a kink in
the $E_{e}-E_{g}$ dependence on $U$ (solid red line) exactly at
$U_{c}$ for and added as well as removed $f$ electron. This is because
below $U_{c}$ the minimal energy requirement always sets $N_{d}=L/2$,
which leads to $N_{d}+N_{f}=L+1$ for a configuration with an additional
$f$ electron and $N_{d}+N_{f}=L-1$ for a removed one. This may seem
strange, considering that $\mu=0$ should ensure the half-filling,
but it is a straightforward consequence of the finite lattice size.
The number of eigenenergies $\varepsilon_{\alpha}$ equals $L$.
These are, for a perfect checkerboard pattern, divided equally into
two bands (Appendix~A). For a disrupted configuration, the subgap
state $\varepsilon_{\mathrm{add}}$ is pulled out from the lower band.
Consequently, as it is energetically advantageous to occupy this state
if $\varepsilon_{\mathrm{add}}<0$, which is the case below $U_{c}$,
this leads to $N_{d}=L/2$. Above $U_{c}$, we have $\varepsilon_{\mathrm{add}}>0$,
which leaves the state unoccupied and therefore $N_{d}=L/2-1$. The
opposite is true for $\varepsilon_{\mathrm{rem}}$. Therefore, the
half-filling for any single of these excitations is restored only
above $U_{c}$ and we need a combination of them to fulfill this requirement
below $U_{c}$. The energy profile for forced condition $N_{f}+N_{d}=L$
is plotted in Figs.~\ref{fig:exc}(d,e) using a dotted line. Note
that the situation for $D=1$ is somewhat more complicated as here
the displacement of a single $f$ particle can have a lower energy
than adding or removing a localized particle at both weak and strong
interaction limits {[}see Fig.~\ref{fig:exc}(e){]}. 
\begin{figure}
\includegraphics[width=1\columnwidth]{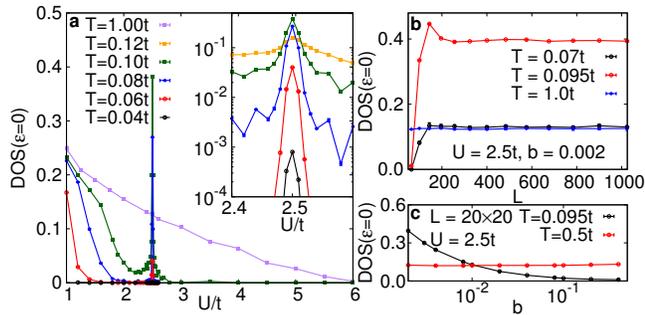}

\caption{(a) Density of states at the Fermi level as a function of $U$ for
a square lattice $L=20\times20$. The inset shows details of the peak
at $U_{c}=2.5t$ where the bound states from Fig.~\ref{fig:exc}(b)
cross each other. (b) Finite size scaling of the DOS at the Fermi
level for various temperatures calculated with broadening $b=0.002t$.
(c) Dependence of the DOS at the Fermi level on the used artificial
broadening. \label{fig:DOS0U}}
\end{figure}

\begin{figure}
\includegraphics[width=1\columnwidth]{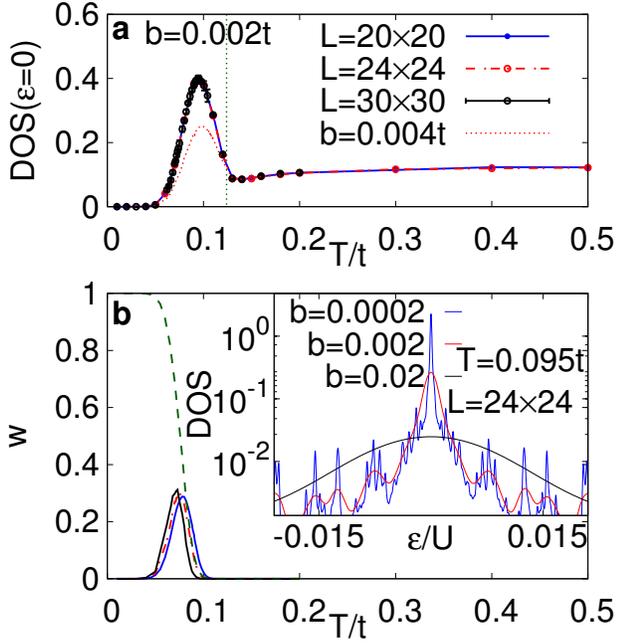}\caption{(a) Density of states at the Fermi level as a function of temperature
for $D=2$ at $U=2.5t$ calculated with broadening parameter $b=0.002t$
(solid lines) for various lattice sizes and with $b=0.004t$ for $L=24\times24$
(dashed line). The vertical black dotted line signals the critical
temperature of the order-disorder phase transition. (b) Dependence
of the statistical weight of the configurations with perfect CDW ordering
(green dashed line calculated for $L=20\times20$) and configurations
with a single $f$ particle added or removed from CDW. \label{fig:DOS0T}}
\end{figure}

The above discussed single $f$-electron excitations have a profound
effect on the DOS even at surprisingly high temperatures. This is
already illustrated by the sharp subgap features in the finite temperature
DOS plotted in Fig.~\ref{fig:DOS}. Nevertheless, we are especially
interested in the ordered phase with high DOS at the Fermi level analogous
to the one observed in infinite dimensions \cite{Lemanski2014}. Therefore,
we show in Fig.~\ref{fig:DOS0U}(a) the dependence of DOS$(\varepsilon=0)$
on $U$ at various temperatures. The highest temperature $T=1t$ represents
the disordered phase, $T=0.12t$ is just above the critical temperature
for $U=2.5t$ and the rest is below it. Figure~\ref{fig:DOS0U}(a)
is a direct $D=2$ analog of the infinite dimensional case shown
in Fig.~7 of the work by Lema\'{n}ski and Ziegler \cite{Lemanski2014}.
Although both cases share some qualitative characteristics, like the
increase of the DOS around $U_{c}$ for small temperatures, there
is one distinct difference. The increase of the DOS around $U_{c}$
is not only much sharper, but for $T\lesssim T_{c}$ it also leads
to values of DOS which are higher than the high temperature limit
where the gap is completely closed. This is again illustrated in Fig.~\ref{fig:DOS0T}(a),
where we show the temperature dependence of the DOS$(\varepsilon=0)$
for $D=2$. The position of the maximum of DOS ($T_{m}\sim0.095t$)
is clearly below the critical temperature ($T_{c}\sim0.12t$). Moreover,
the DOS profile is stable with respect to the lattice size as it is
illustrated in Fig.~\ref{fig:DOS0T}(a), Fig.~\ref{fig:DOS0U}(b),
and in the inset of Fig.~\ref{fig:DOS}(b), where we show the detail
of the DOS around the Fermi level.

The above-discussed DOS were calculated with an artificial Gaussian
broadening of the $\delta$-functions with $b=0.002t$. The effect
of the Gaussian broadening on the DOS$(\varepsilon=0)$ is completely
negligible in the disordered phase as illustrated in Fig.~\ref{fig:DOS0U}(c)
by the red circles. The situation in the ordered CDW phase is more
complicated, especially for critical $U_{c}$ and low temperatures.
Figure~\ref{fig:DOS0U}(c) shows that the DOS$(\varepsilon=0)$ calculated
at $T=0.095t$ (black circles) increases with the decreasing broadening.
Also in Fig.~\ref{fig:DOS0T}(a), one can see by following the dashed
line, which represents the DOS$(\varepsilon=0)$ calculated for $L=20\times20$
and broader broadening of $b=0.004t$, that bigger artificial broadening
leads to a decrease of the calculated DOS$(\varepsilon=0)$ maximum.

The reason is twofold. For the $D=2$, the band around $\varepsilon=0$
has a fine structure, as is shown in the inset of Fig.~\ref{fig:DOS}(b),
and has a clear maximum at $\varepsilon=0$. A wide artificial broadening
smooths this structures, which significantly lowers the DOS at the
Fermi level. From our data it is not clear if this fine structure
will disappear in the thermodynamic limit. It seems not to be the
case, as the detail of the DOS shown in the inset of Fig.~\ref{fig:DOS}(b)
depends only weakly on the lattice size, but a study on much bigger
lattices is required to confirm this.

Even more important is that despite the broadening into a band provided
by multi-particle excitations, the states at the Fermi level stay
very sharp even for large lattices, as is shown in the inset of
Fig.~\ref{fig:DOS0T}(b). This effect becomes even more crucial with
decreasing temperature as here the configurations with just one $f$
particle added or removed from a perfect checkerboard ordering can
have a very high weight. This is shown in Fig.~\ref{fig:DOS0T}(b),
where we plotted the total MC weight of these configuration $w\sim\left(e^{-\beta F(w_{\mathrm{add}})}+e^{-\beta F(w_{\mathrm{rem}})}\right)/Z$
as a function of temperature for various lattice sizes (solid lines)
and compare it with the analogous weight for checkerboard ordering
(dashed line). There is a clear maximum at which the combined weight
of the above excited states is almost one-third of the total one.
This can explain the extremely sharp subgap states for low temperatures
in the finite system shown in Fig.~\ref{fig:DOS} \footnote{Some caution is in place regarding this statement. The used MC is
of a single-flip character which could in principle distort the statistics
if there would be some low energy $f$-electron configurations, which
are hard to reach by a series of single particle updates from e.g.
the checkerboard ordering. Nevertheless, we have tested this by using 
a big set of random initial conditions and this does
not seem to be the case.}.

\begin{figure}
\includegraphics[width=1\columnwidth]{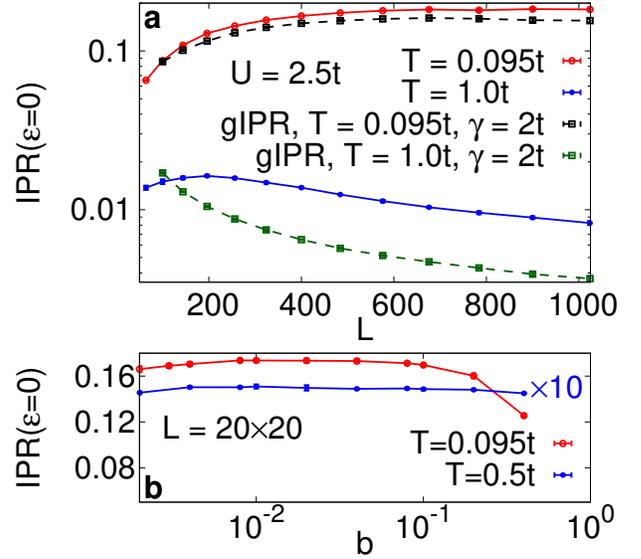}\caption{(a) System size scaling of the averaged inverse participation ratio
for $U=2.5t$ at $T=0.095t$ (red circles) and $T=1t$ (blue circles)
and system size scaling of the generalized averaged inverse participation
ratio for the same parameters and $\gamma=2t$ (black and green squares).
(b) Dependence of the inverse participation on the artificial broadening
of the delta functions. The IPR for $T=0.5t$ is multiplied by factor
of $10$.\label{fig:IPR}}
\end{figure}

The enhanced DOS at the Fermi level caused by the crossing of the
sharp maxima might raise a question if the phase has still an insulating
character. However, knowing that the prevailing mechanism behind this
effect is the impurity like single-particle excitation of the localized
subsystem, one can expect a strong localization of the $d$ electrons.
We show by studying the scaling of the averaged IPR($\varepsilon=0$)
depicted in Fig.~\ref{fig:IPR}(a) that this is really the case.
For localized states, the IPR should saturate with increasing lattice
size to a finite value. Note that IPR$\rightarrow1$ would point to
a complete localization of the $d$ electrons to a single lattice
point. Therefore, the saturation of IPR to $\sim0.16$ for $T=0.095t$
shown in Fig.~\ref{fig:IPR}(a) (red circles) still points to a strong
localization. On the other hand, the slow decline of the IPR for $T=1t$
(blue circles) points to a weak localization as expected in this regime
for a finite size system \cite{Antipov2016,Zonda2018}. 

As shown in Fig.~\ref{fig:IPR}(b), the IPR is, in contrast to the
DOS results discussed above, relatively stable for a broad range of
broadening parameter $b$ (note the logarithmic scale) even below
the critical temperature. Therefore, the conclusions of strong localization
of the crossed states is not affected by the artificial broadening
of the $\delta$-functions.

To further support these findings, we provide an alternative test
of the above conclusions in the next subsection. Instead of artificial
broadening, we consider a heterostructure where the system is coupled
to two semi-infinite metallic leads. These provide a different kind
of broadening of the system states. It can be argued that this broadening
is more natural, but it is also uneven, because the broadening of
the LDOS decreases with increasing distance from the system-leads
interface \cite{Freericks2004,Zonda2018}. In addition, this setup
allows us to study the transport thought a finite system.

\subsection{Transport properties of a heterostructure\label{subsec:Transport}}

\begin{figure}
\includegraphics[width=1\columnwidth]{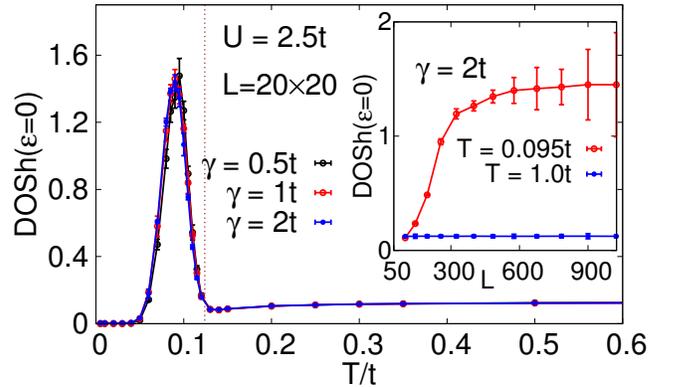}

\caption{Averaged DOS of the coupled system calculated for $U=2.5t$, $L=20\times20$
and three values of $\gamma$. The position of the maximum $T\sim0.095t$
coincides with the position of the maxima in Fig.~\ref{fig:DOS0T}(a).
The inset represents the finite size scaling of DOS at $T=0.095t$
(maximum) and $1t$ (high temperature). \label{fig:DOSh}}
\end{figure}

We have shown recently \cite{Zonda2018} that the typical phases of
the FKM have different influence on the transport properties of a
heterostructure. However, that study did not focus on the particular
case where the system is in CDW phase but has a large density of states
at the Fermi level as is the case in Fig.~\ref{fig:DOS0T}(a).
Here we study the effect of the finite DOS in this regime on the transport
properties for a finite dimensional central system. Because of that,
we first have to readdress the problem of DOS and localization for
the heterostructure. 
\begin{figure}
\includegraphics[width=0.95\columnwidth]{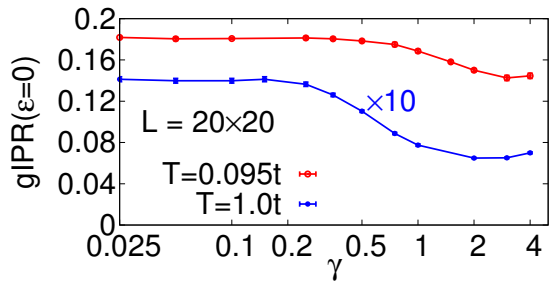}\caption{gIPR as a function of $\gamma$ for $T=0.095t$ and $T=1t$ and central
system size $L=20\times20$. The gIPR for $T=1t$ is multiplied by
factor of $10$ for the sake of clarity.\label{fig:gIPRg}}
\end{figure}

In contrast to the isolated FKM studied in the previous section, in
the heterostructure the broadening of the system states results naturally
from the coupling to the semi-infinite leads. In Fig.~\ref{fig:DOSh},
we show the averaged DOSh$(\varepsilon=0)$ for $L=20\times20$ and
three values of $\gamma$. The DOSh$(\varepsilon=0)$ above the critical
temperature, signaled by vertical dotted line, is identical to the
one in Fig.~\ref{fig:DOS0T}(a) and does not depend on the lattice
size as can be seen in the inset of Fig.~\ref{fig:DOSh} (blue line).
The positions of the maxima are in compliance as well ($T\sim0.095t$)
and the maximum of DOSh$(\varepsilon=0)$ is well above its value
in the disordered phase. In addition, the DOSh$(\varepsilon=0)$ shown
in Fig.~\ref{fig:DOSh} depends only weakly on the chosen values
of $\gamma$'s. This supports the conclusion that the crossing of
the major subgap bands at critical $U$ can lead to a DOS at the
Fermi level, which exceeds its values in the gapless disordered phase.
On the other hand, the DOSh$(\varepsilon=0)$ calculated for $T=0.095t$
depends much stronger on the lattice size (red circles in the inset
of Fig.~\ref{fig:DOSh}) than the equivalent DOS calculated for isolated
FKM. This, as well as the increasing error bars, is a direct consequence
of the fact that the broadening coming from the leads is not homogeneous
in the central system as its effect on the LDOSh decreases with the
distance from the system-leads interfaces \cite{Zonda2018}. Consequently,
the broadening in the central part of the system coming from the leads
might vanish fast with the increased lattice size. 

The qualitative differences between the natural and artificial broadening
allow us to perform an alternative investigation of the localization
based on the gIPR defined in Eq.~\eqref{eq:gIPR}. A direct comparison
of the finite size scaling of the gIPR calculated for $\gamma=2t$
with the IPR is shown Fig.~\ref{fig:IPR}(a). The gIPR for $T=0.095t$
(black squares) has the same profile as IPR and it convergences to
a similar finite value, which confirms strong localization even for
the coupled system. Similarly, the scaling of the gIPR in the disordered
phase represented by $T=1t$ (green squares) points to a weakly localized
central system at best (see also Ref.~\cite{Zonda2018}). The comparison
with the IPR also reveals that the coupling to the leads can suppress
the localization in the finite system in this regime.

We analyze the effect of coupling to the leads on the localization
in the finite system ($L=20\times20$) in more detail in Fig.~\ref{fig:gIPRg}.
Both gIPR curves calculated at $T=0.095t$ and $T=1t$ are saturated
at low values of the coupling $\gamma$. The saturated values are
in good agreement with the ones calculated for the isolated system
{[}Fig.~\ref{fig:IPR}(b){]}. This can also be interpreted as
evidence that the coupling to the leads provides a good independent
method for studying the problem of the localization.

However, as we increase $\gamma$ (note the logarithmic scale in Fig.~\ref{fig:gIPRg})
the gIPR significantly decreases for $\gamma\apprge0.3t$ at $T=1t$
and $\gamma\apprge1t$ for $T=0.095t$. This effect is stronger in
the disordered phase, where at strong coupling the already small gIPR
drops to half its weak coupling value. Nevertheless, the localization
is weakened in the ordered phase as well. It is therefore worth it to
examine how the coupling to the leads affects the transport properties
in a finite system.

\begin{figure}
\includegraphics[width=0.95\columnwidth]{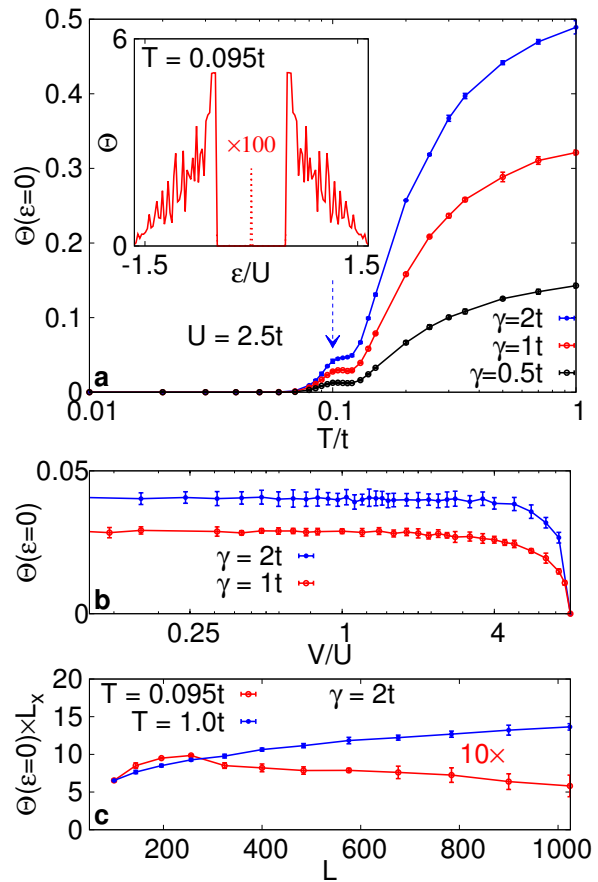}\caption{(a) Dependence of the equilibrium transmission function on temperature
calculated at Fermi level for a heterostructure with $U=2.5t$ and
the system size $L=20\times20$ coupled to two semi-infinite noninteracting
leads with semi-elliptical surface density of states and band half-width
$B=20t$. The position of the humps signalled by an arrow coincides
with the maxima in Fig.~\ref{fig:DOS0T}. The inset is an example
of the full transmission function for $\gamma=1t$ and $T=0.095t$.
The subgap region is multiplied by $100$ for the sake of visibility.
(b) Nonequilibrium transmission function for $\varepsilon=0$ as a
function of voltage drop rescaled by $U=2.5t$. (c) Dependence of
the equilibrium transmission function at the Fermi level multiplied
by the linear size of the system on the total lattice size. The values
for $T=0.095t$ were scaled by factor $10$. \label{fig:TF}}
\end{figure}

We focus on the transmission function as this provides the most detail
information on the charge transport. Figure~\ref{fig:TF}(a) shows
the transmission function at $\varepsilon=0$ as a function of temperature
for $U=2.5t$, central system size $L=20\times20$ and for three values
of system lead hopping $\gamma$. We focus on the equilibrium situation
($V=0$) because a voltage that is smaller than the CDW gap ($V<U/2$)
has only a small effect on the transmission function at $\varepsilon=0$.
This is shown in Fig.~\ref{fig:TF}(b) for $U=2.5t$ and $T=0.095t$,
where voltage values are spread on a logarithmic scale.

The transmission function in Fig.~\ref{fig:TF}(a) is negligible
for low temperatures, but a clear ``hump'' for $\gamma=2t$ and
local maximum for $\gamma<2t$ (signaled by an arrow) are present
close to the temperature where the DOS of the central system has its
maximum {[}see Fig.~\ref{fig:DOSh}{]}. The transmission function
at Fermi level is still several magnitudes smaller than its values
outside the CDW gap, as seen in the inset of Fig.~\ref{fig:TF}(a),
and it is negligible for any energy within the gap with the exception
of the close vicinity of $\varepsilon=0$. Nevertheless, its clear
increase close to $T\sim0.1t$ shows that, despite the relatively
strong localization, the crossing of the subgap states can influence
the charge transport through a finite system. Interestingly, it actually
enables a finite charge transmission otherwise blocked by the CDW
gap. Still, the transmission function significantly increases above
the critical temperature of the order-disorder transition and, despite
much lower DOS in this phase, the transmission function can be several
times higher than at $T\sim0.1t$.

This clearly reflects the difference of the localization in the ordered
and disordered phases already shown in Fig.~\ref{fig:IPR}(a). This
effect can also be seen in the system size scaling of the transmission
functions plotted in Fig.~\ref{fig:TF}(c). To highlight the difference,
we scaled the transmission function by the linear size of the system.
The weak localization in the disordered phase ($T=1t$), further
lowered by the coupling to the leads, results in a situation where
the scaled transmission function increases with $L$. On the other
hand, the scaled transmission for $T=0.095t$ , which even for small
system lattices is approximately ten times smaller than for $T=1t$,
rapidly decreases with growing lattice size. The strong localization
in this regime clearly overrules even the increasing DOSh($\varepsilon=0$)
shown in the inset of Fig.~\ref{fig:DOSh}.

We can therefore conclude that the high density of states in the CDW
phase observed in the vicinity of $U_{c}$ can lead to a notable increase
of charge transport through a finite system, but this effect rapidly
vanishes with increasing system size. The reason is the relatively
strong localization of the states at the Fermi level.

\section{Conclusions\label{sec:Conclusions}}

We have studied in realistic dimensions the structure of the subgap
states of the FKM in the ordered CDW phase. We have shown that, similar
to exact results in infinite dimensions, there are pronounced maxima
in the subgap DOS placed symmetrically around the Fermi level which
merge around some critical value of $U$ and exchange their positions
above it. The position of these maxima does not depend on the temperature,
because they mainly reflect the underlying sharp states resulting
from single-particle excitations of the localized $f$-electron subsystem.
The most pronounced of these states belong to an addition and removal
of a single $f$ electron from a perfect checkerboard ordering.

The crossing of the most pronounced subgap maxima leads to a rise
of the DOS at the Fermi level which can exceed even the DOS in the
disordered gapless phase. However, the states in this regime are strongly
localized. We have confirmed this by studying DOS and IPR for two
different setups. First, we addressed an isolated FKM, where we used
an artificial broadening of the states. Second, we studied a heterostructure,
where the states had been broadened by coupling the system to simple
semi-infinite leads. Although the coupling to the leads can lower
the localization in the disordered phase, both studies have lead to
the same qualitative conclusions.

In the case of the heterostructure, we have also shown that the significant
increase of the DOS at critical $U$ can boost the charge transport
trough a small finite system in the ordered phase. It increases the
charge transmission which is otherwise suppressed in the whole range
of energies within the CDW gap. Here, the strength of the coupling
to the leads plays an important role. Nevertheless, with increasing
lattice size this effect is quickly suppressed by the strong localization. 
\begin{acknowledgments}
The authors acknowledge support by the state of Baden-W\"urttemberg
through bwHPC and the German Research Foundation (DFG) through Grant
No. INST 40/467-1 FUGG. J. O. acknowledges support from Georg H. Endress
Foundation. We thank Romuald Lema\'{n}ski and James K. Freericks for introducing
us to the problem of the crossing of the subgap states in FKM and
Tom\'{a}\v{s} Novotn\'y for very helpful discussions. 
\end{acknowledgments}

\section*{Appendix A: CDW gap\label{sec:Appendix-A:-CDW}}

Here we present the ground-state solution of the spinless FKM on a
hypercubic lattice. We show that the checkerboard ordering of the
$f$ electrons opens a gap in the DOS and derive general formulas
for eigenvalues and eigenvectors. 

The Hamiltonian Eq.~\eqref{eq:Model} for the CDW ordering of $f$ electrons
in hyper cubic lattice reads

\begin{eqnarray}
H_{\mathrm{FK}}^{\mathrm{CDW}} & = & -t\sum_{\boldsymbol{\mathbf{l}},\boldsymbol{\boldsymbol{\delta}}}\left(d_{\boldsymbol{\mathbf{l}}}^{\dagger}d_{\mathbf{\boldsymbol{l}}+\boldsymbol{\boldsymbol{\delta}}}^{\phantom{\dagger}}+h.c.\right)+\frac{U}{2}\sum_{\mathbf{\boldsymbol{l}}}e^{i\boldsymbol{\pi}\mathbf{\boldsymbol{l}}}d_{\mathbf{\boldsymbol{l}}}^{\dagger}d_{\boldsymbol{\mathbf{l}}}^{\phantom{\dagger}},\label{eq:CDW}
\end{eqnarray}
where $\boldsymbol{\delta}$ indices's the relevant nearest neighbors
(e.g., $1\boldsymbol{i}_{x}+0\boldsymbol{i}_{y}+0\boldsymbol{i}_{z}$).
Fourier transformation, given by 
\begin{equation}
d_{\boldsymbol{k}}^{\dagger}=\frac{1}{\sqrt{L}}\sum_{\mathbf{\boldsymbol{l}}}e^{i\mathbf{\boldsymbol{k}\cdot\boldsymbol{l}}}d_{\mathbf{\boldsymbol{l}}}^{\dagger},\label{eq:dk}
\end{equation}
leads to 
\begin{equation}
H_{\mathrm{FK}}^{\mathrm{CDW}}=\sum_{\boldsymbol{\boldsymbol{k}}}\epsilon_{\boldsymbol{k}}d_{\boldsymbol{\boldsymbol{k}}}^{\dagger}d_{\mathbf{\boldsymbol{k}}}^{\phantom{\dagger}}+\frac{U}{2}\sum_{\mathbf{\boldsymbol{k}}}d_{\mathbf{\boldsymbol{k}}}^{\dagger}d_{\boldsymbol{\boldsymbol{k}+\boldsymbol{\pi}}}^{\phantom{\dagger}},\label{eq:Hch}
\end{equation}
where

\[
\epsilon_{\boldsymbol{k}}=-2t\sum_{i=1}^{D}\cos\left(k_{i}\right).
\]
By replacing of the summation by integration and careful reshaping
of the Brillouin zone, the Hamiltonian in Eq.~\eqref{eq:Hch} can
be written in a matrix form 
\begin{eqnarray}
H_{\mathrm{FK}}^{CDW} & = & \prod_{i=2}^{D}\intop_{-\pi}^{\pi}\frac{dk_{i}}{2\pi}\intop_{-\frac{\pi}{2}}^{\frac{\pi}{2}}\frac{dk_{1}}{2\pi}\label{eq:kHam}\\
 &  & \left(\begin{array}{cc}
d_{\boldsymbol{k}}^{\dagger} & d_{\boldsymbol{k}+\boldsymbol{\pi}}^{\dagger}\end{array}\right)\left(\begin{array}{cc}
\epsilon_{\boldsymbol{k}} & \frac{U}{2}\\
\frac{U}{2} & \epsilon_{\boldsymbol{k}+\boldsymbol{\pi}}
\end{array}\right)\left(\begin{array}{c}
d_{\boldsymbol{k}}^{\phantom{\dagger}}\\
d_{\boldsymbol{k}+\boldsymbol{\pi}}^{\phantom{\dagger}}
\end{array}\right).\nonumber 
\end{eqnarray}
Therefore, the states are divided into two bands separated by CDW
gap with width $U$ 
\[
\varepsilon_{\boldsymbol{k}}^{\pm}=\pm\sqrt{\epsilon_{\boldsymbol{k}}^{2}+\left(\frac{U}{2}\right)^{2}}.
\]
The related eigenvectors stored as a columns in matrix $Q_{\boldsymbol{k}}$
read 
\begin{equation}
Q_{\boldsymbol{k}}=\left(\begin{array}{cc}
\frac{\epsilon_{\mathbf{\boldsymbol{k}}}+\varepsilon_{\mathbf{\boldsymbol{k}}}^{+}}{\sqrt{\left(\frac{U}{2}\right)^{2}+\left(\epsilon_{\mathbf{\boldsymbol{k}}}+\varepsilon_{\mathbf{\boldsymbol{k}}}^{+}\right)^{2}}}, & \frac{\epsilon_{\mathbf{\boldsymbol{k}}}+\varepsilon_{\mathbf{\boldsymbol{k}}}^{-}}{\sqrt{\left(\frac{U}{2}\right)^{2}+\left(\epsilon_{\mathbf{\boldsymbol{k}}}+\varepsilon_{\mathbf{\boldsymbol{k}}}^{-}\right)^{2}}}\\
\frac{U/2}{\sqrt{\left(\frac{U}{2}\right)^{2}+\left(\epsilon_{\mathbf{\boldsymbol{k}}}+\varepsilon_{\mathbf{\boldsymbol{k}}}^{+}\right)^{2}}}, & \frac{U/2}{\sqrt{\left(\frac{U}{2}\right)^{2}+\left(\epsilon_{\mathbf{\boldsymbol{k}}}+\varepsilon_{\mathbf{\boldsymbol{k}}}^{-}\right)^{2}}}
\end{array}\right).
\end{equation}

\section*{Appendix B: Subgap states\label{sec:Appendix-B}}

In this Appendix, some analytical results for the positions of the
subgap states resulting from single $f$-electron excitations are
derived.

Let us excite the groundstate Hamiltonian in Eq.~\eqref{eq:Hch}
by adding or removing a single $f$-electron on position $\boldsymbol{m}$
\[
H_{\mathrm{ex}}=-Ue^{i\boldsymbol{\pi}\boldsymbol{m}}d_{\boldsymbol{m}}^{\dagger}d_{\boldsymbol{m}}^{\phantom{\dagger}}.
\]
Our aim here is to find the position of the sharp states in the spectra
resulting from this single impurity. The Hamiltonian (\ref{eq:CDW})
can be formally diagonalized by unitary transformation, 
\begin{eqnarray*}
\widetilde{d}_{\mathbf{\alpha}}^{\dagger} & = & \sum_{\boldsymbol{l}}\psi_{\alpha\boldsymbol{l}}d_{\boldsymbol{l}}^{\dagger},\\
\widetilde{d}_{\mathbf{\alpha}}^{\phantom{\dagger}} & = & \sum_{\boldsymbol{l}}d_{\boldsymbol{l}}\psi_{\boldsymbol{l}\alpha}^{\dagger},
\end{eqnarray*}
after which the total system Hamiltonian including $H_{\mathrm{ex}}$
reads 
\[
H=\sum_{\alpha}\varepsilon_{\alpha}\widetilde{d}_{\alpha}^{\dagger}\widetilde{d}_{\alpha}^{\phantom{\dagger}}-Ue^{i\boldsymbol{\pi}\boldsymbol{m}}\sum_{\alpha,\beta}\psi_{\alpha\boldsymbol{m}}\psi_{\boldsymbol{m}\beta}^{\dagger}\widetilde{d}_{\alpha}^{\dagger}\widetilde{d}_{\beta}^{\phantom{\dagger}}.
\]
We define the Green's function, 
\[
G_{\alpha,\beta}(\tau)=-i\Theta(\tau)\left\langle \left[\widetilde{d}_{\alpha}^{\phantom{\dagger}}(\tau),\widetilde{d}_{\beta}^{\dagger}(0)\right]_{+}\right\rangle 
\]
and use the equation of motion technique to get 
\begin{eqnarray*}
\left(i\frac{d}{d\tau}-\varepsilon_{\alpha}\right)G_{\alpha,\beta}(\tau) & = & \delta_{\alpha,\beta}\delta(\tau)\\
 &  & -Ue^{i\boldsymbol{\pi m}}\sum_{\alpha'}\psi_{\alpha\boldsymbol{m}}\psi_{\boldsymbol{m}\alpha'}^{\dagger}G_{\alpha',\beta}(\tau).
\end{eqnarray*}
After the transformation into energy domain this reads 
\[
\left(\varepsilon-\varepsilon_{\alpha}+i0\right)G_{\alpha,\beta}(\varepsilon)=\delta_{\alpha,\beta}-Ue^{i\boldsymbol{\pi m}}\sum_{\alpha'}\psi_{\alpha\boldsymbol{m}}\psi_{\boldsymbol{m}\alpha'}^{\dagger}G_{\alpha',\beta}(\varepsilon),
\]
or written in matrix form 
\[
\left(\mathbf{E}+Ue^{i\pi\boldsymbol{m}}\boldsymbol{\psi^{m}}\right)\mathbf{G}(\varepsilon)=\mathbf{1},
\]
where $\mathbf{E}$ is diagonal matrix with elements $E_{\alpha,\alpha}=\varepsilon-\varepsilon_{\alpha}+i0$.
We are primarily interested in the position of the bound states inside
the CDW gap which can be calculated from the zeros of the determinant
of the inverse Green's function. Because $\mathbf{E}$ is diagonal
and $\psi_{\alpha\alpha'}^{\boldsymbol{m}}=\psi_{\alpha\boldsymbol{m}}\psi_{\boldsymbol{m}\alpha'}^{\dagger}$
we can use Sylvester's determinant theorem {[}$\det\left(X+cr\right)=\det(X)(1+rX^{-1}c)${]}:

\begin{eqnarray}
\det\left(\mathbf{G}^{-1}(\varepsilon)\right) & = & \left(1+Ue^{i\boldsymbol{\pi}\boldsymbol{m}}\sum_{\alpha}\frac{\psi_{\boldsymbol{m}\alpha}^{\dagger}\psi_{\alpha\boldsymbol{m}}}{\varepsilon-\varepsilon_{\alpha}+i0}\right)\nonumber \\
 &  & \times\prod_{\alpha}\left(\varepsilon-\varepsilon_{\alpha}+i0\right).\label{eq:det}
\end{eqnarray}
The position of the sharp states in the CDW gap can therefore be obtained
by solving 
\begin{equation}
1+e^{i\boldsymbol{\pi}\boldsymbol{m}}U\sum_{\alpha}\frac{\psi_{\boldsymbol{m}\alpha}^{\dagger}\psi_{\alpha\boldsymbol{m}}}{\varepsilon-\varepsilon_{\alpha}}=0.\label{eq:sol}
\end{equation}
We evaluate this equation by using the transformation from Eq.~\eqref{eq:dk}
together with the decomposition $M=Q\Lambda Q^{-1}$, where $M$ is
the inner matrix from Eq.~\eqref{eq:kHam} and elements of the diagonal
matrix $\Lambda$ are $\varepsilon_{\boldsymbol{k}}^{+}$ and $\varepsilon_{\boldsymbol{k}}^{-}$.
The product in Eq.~\eqref{eq:sol} then reads 
\[
\psi_{\boldsymbol{m}\alpha}^{\dagger}\psi_{\alpha\boldsymbol{m}}=\frac{1}{L}\sum_{p=\pm}\left(1+Ue^{-i\boldsymbol{\pi}\boldsymbol{m}}\frac{\varepsilon_{\boldsymbol{k}}^{p}+\epsilon_{\boldsymbol{k}}}{\left(\frac{U}{2}\right)^{2}+\left(\varepsilon_{\boldsymbol{k}}^{p}+\epsilon_{\boldsymbol{k}}\right)^{2}}\right),
\]
and the sum over index $\alpha$ changes to 
\[
\sum_{\alpha}\rightarrow\prod_{i=2}^{D}\intop_{-\pi}^{\pi}\frac{dk_{i}}{2\pi}\intop_{-\frac{\pi}{2}}^{\frac{\pi}{2}}\frac{dk_{1}}{2\pi}\equiv\sum_{\mathbf{K}}.
\]
Using the new notation, Eq.~\eqref{eq:sol} can be rewritten to a
simple form 
\begin{equation}
1=e^{i\boldsymbol{\pi}\boldsymbol{m}}\frac{U}{L}\sum_{\mathbf{K}}\frac{Ue^{-i\boldsymbol{\pi}\boldsymbol{m}}+2\varepsilon}{\varepsilon_{\boldsymbol{k}}^{2}-\varepsilon^{2}}.\label{eq:sol2}
\end{equation}
In $D=1$ this equation can be easily evaluated and reads 
\[
\frac{2\left(U+2\varepsilon e^{i\pi m}\right)U}{\sqrt{U^{2}-4\varepsilon^{2}}\sqrt{16t^{2}+U^{2}-4\varepsilon^{2}}}=1.
\]
Therefore, the solution for $U$ as a function of the energy of bound
states is 
\[
U=4\sqrt{\frac{\pm1-2\tilde{\varepsilon}}{\left(\pm3-2\tilde{\varepsilon}\right)(1\pm2\tilde{\varepsilon})^{2}}},
\]
where $\tilde{\varepsilon}=\varepsilon/U$. At the crossing of the
states, we have $\varepsilon=0$ and therefore the critical interaction
in $D=1$ is $U_{c}=4/\sqrt{3}$.

The evaluation of Eq.~\eqref{eq:sol2} in $D=2$ leads to 
\[
1=\frac{U^{2}(1\pm2\tilde{\varepsilon})}{\pi u}\sqrt{\frac{u}{16+u}}\mathcal{K}\left(\frac{16}{16+u}\right),
\]
where $u=U^{2}(0.25-\tilde{\varepsilon}^{2})$ and $\mathcal{K}(x)$
is the elliptic integral of the first kind. The critical interaction
in $D=2$ is $U_{c}\doteq2.5t$ (with rounding at the third decimal
place). The $D=3$ case can be evaluated numerically and has the critical
interaction $U_{c}\cong3.18t$.

The displacement of the $f$ electron to a neighboring position can
be seen as adding and removing an $f$ electron at adjoining positions
and can be represented by an additional term, 
\[
H_{\mathrm{ex}}=-Ue^{i\boldsymbol{\pi}\boldsymbol{m}}(d_{\boldsymbol{m}}^{\dagger}d_{\boldsymbol{m}}^{\phantom{\dagger}}-d_{\boldsymbol{m}+1i_{n}}^{\dagger}d_{\boldsymbol{m}+1i_{n}}^{\phantom{\dagger}}),
\]
where $i_{n}$ choses the direction of the $f$-electron shift. This
leads to a more complicated formula for the determinant 
\begin{widetext}
\begin{eqnarray*}
\det\left(\mathbf{G}^{-1}(\varepsilon)\right) & = & \prod_{\alpha}\left(\varepsilon-\varepsilon_{\alpha}+i0\right)\left[\left(1+Ue^{i\boldsymbol{\pi m}}\sum_{\alpha}\frac{\psi_{\boldsymbol{m}\alpha}^{\dagger}\psi_{\alpha\boldsymbol{m}}}{\varepsilon-\varepsilon_{\alpha}+i0}\right)\left(1-Ue^{i\boldsymbol{\pi m}}\sum_{\alpha}\frac{\psi_{\boldsymbol{m}+1i_{n},\alpha}^{\dagger}\psi_{\alpha\boldsymbol{m}+1i_{n}}}{\varepsilon-\varepsilon_{\alpha}+i0}\right)\right.\\
 &  & +\left.\left(Ue^{i\boldsymbol{\pi m}}\sum_{\alpha}\frac{\psi_{\boldsymbol{m}+1_{n},\alpha}^{\dagger}\psi_{\alpha\boldsymbol{m}}}{\varepsilon-\varepsilon_{\alpha}+i0}\right)\left(Ue^{i\boldsymbol{\pi m}}\sum_{\alpha}\frac{\psi_{\boldsymbol{m},\alpha}^{\dagger}\psi_{\alpha\boldsymbol{m}+1i_{n}}}{\varepsilon-\varepsilon_{\alpha}+i0}\right)\right].
\end{eqnarray*}
The sums in this formula can be evaluated analogously to the previous
case, but their evaluation leads to much more complicated expressions,
which we therefore omit here. 
\end{widetext}


%

\end{document}